\def\q{\hfill\rule{0.5ex}{1.3ex}}
\documentclass[amstex,12pt]{article} % OL 11pt is better
\usepackage{amssymb,amsmath,amsthm}
\usepackage{graphics,color}
\usepackage{tikz}
\title{A note on the dynamic dominant resource fairness mechanism} % in cloud computing systems
\author{Weidong Li% \thanks{Corresponding author. \newline E-mail addresses: weidong@ynu.edu.cn, lxghost@126.com, zxl@ynu.edu.cn, xjzhang@ynu.edu.cn}
, Xi Liu,  Xiaolu Zhang, Xuejie Zhang
 \\
Yunnan University, Kunming, 650091, PR China\\
E-mail: \{weidong,zxl,xjzhang\}@ynu.edu.cn, lxghost@126.com}
\begin{document}

\maketitle

\noindent \textbf{Abstract.} Multi-resource fair allocation has been
a hot topic of resource allocation. Most recently,   a dynamic {\it
dominant resource fairness} (DRF) mechanism is proposed for dynamic
multi-resource fair allocation.  In this paper, we prove that the
competitive ratio  of the dynamic DRF mechanism is the reciprocal of
the number of resource types, for two different objectives.
Moreover, we
develop a linear-time algorithm to find a dynamic DRF solution at each step. 
\section{Introduction}
 With the ever-growing demand for cloud resources, multi-resource
(such as CPUs, memory, and bandwidth) fair allocation became a
fundamental problem in cloud computing systems. The traditional
slot-based scheduler for state-of-the-art  cloud computing
frameworks (for example, Hapdoop) can lead to poor performance,
unfairly punishing certain workloads. Ghodsi et al. \cite{Ghodsi11}
  proposed a compelling alternative known as the {\it
dominant resource fairness} (DRF) mechanism, which is designed for
{\it Leontief } preferences. DRF is to maximize the minimum dominant
share of users, where the dominant share is the maximum share of any
resource allocated to that user.
 DRF is generally applicable to multi-resource environments where users have
heterogeneous demands, and is now implemented in the Hadoop Next
Generation Fair Scheduler.

In recent years, DRF has attracted much attention and been
generalized to many dimensions.  Joe-Wong et al. \cite{Wong13}
designed  a unifying multi-resource allocation framework that
captures the trade-offs between fairness and efficiency, which
generalizes the DRF measure. Gutman and Nisan \cite{Gutman12}
situated DRF  in a common
 economics framework, obtaining a general economic perspective.
% Bhattacharya et al. \cite{Bhattacharya13} generalized
%DRF to a hierarchical scheduler that offers service isolations in a
%computing system with a hierarchical structure.
 Parkes et al.
\cite{Parkes15} extended DRF in several ways, including the presence
of zero demands and the case of indivisible tasks.
%Li et al.\cite{Li14} extended DRF to the finite case, where the number of
%t asks for each user is bounded.
Wang et al. \cite{Wang14}
generalized the DRF measure into the cloud computing systems with
heterogeneous servers.
%Psomans and Schwartz \cite{Psomas13}, and
%Friedman, Ghodsi, and Psomas \cite{Friedman14}
% studied the multi-resource allocation of discrete tasks on multiple machines.
Most recently, Zarchy, Hay and Schapira \cite{Zarchy}
developed a framework for fair resource allocation that captures
such implementation tradeoffs by allowing users to submit multiple
resource demands.

DRF uses complete information about the requirements of all agents
in order to find the fair solution. However, in reality, agents
arrive over time, and we do not know the requirements of forthcoming
agents before allocating the resources to the arrived agents.
Recently, Kash, Procaccia and Shah \cite{Kash14} introduced a
dynamic model of fair  allocation and proposed a dynamic DRF
mechanism. They mentioned
 that a dynamic DRF solution can be found by using water-filling algorithm or solving the
 corresponding linear program. However, the running time of the
water-filling algorithm is pseudo-polynomial in worst-case scenario.
Although solving a linear program  can be done within polynomial
time, the running time is high. It is desired to design an efficient
algorithm to find a dynamic DRF solution.

In this paper, we further study the dynamic DRF mechanism.  The rest
of the paper is organized as follows. Section 2 describes the
dynamic DRF mechanism.  Section 3 gives the competitive ratios
analysis
 of the dynamic DRF mechanism.  Section 4 presents a
polynomial-time algorithm, which can find a dynamic DRF solution in
$O(k)$
 time at every step $k$. Finally,
Section 5 concludes the paper and gives the future work.

\section{Dynamic Dominant Resource Fairness}
Throughout this paper, assume that resources are divisible. In a
multi-resource environment, there are $n$ agents and $m$ resources.
Each agent $i$ requires $D_{ir}$-fraction of resource $r$ for each
task, assuming that $D_{ir}>0$ for each resource $r$. As defined in
\cite{Ghodsi11}, the {\it dominant resource} of agent $i$ is the
resource $r^*_i$ such that
 $D_{ir^*_i}=\max_{r}D_{ir}$, and $D_{ir^*_i}$ is called its {\it dominant share}. Following
\cite{Kash14,Parkes15}, the {\it normalized  demand vector} of agent
$i$ is given by $\textbf{d}_i=(d_{i1},\ldots,d_{im})$, where
$d_{ir}=D_{ir}/D_{ir^*_i}$ for each resource $r=1,\ldots,m$.
Clearly, $d_{ir}\leq 1$ and $d_{ir^*_i}=1$ for each agent $i$.

In the dynamic resource allocation model considered in
\cite{Kash14}, agents arrive at different times and do not depart.
Assume that agent 1 arrives first, and in general agent $k$ arrives
after agents $1$, $\ldots$, $k-1$, for $k\geq 2$. For convenience,
we say that agent $k$ arrives in {\it step} $k$. An agent reports
its demand  which does not change over time when it arrives. Thus,
at step $k$, demand vectors $\textbf{d}_1$, $\ldots$, $\textbf{d}_k$
are known, and demand vectors  $\textbf{d}_{k+1}$, $\ldots$,
$\textbf{d}_n$ are unknown. At each step $k$, a dynamic DRF
mechanism produces an allocation $\textbf{A}^k$ over the agents
present in the system, where $\textbf{A}^k$ allocates
$A^k_{ir}$-fraction of resource $r$ to agent $i$, subject to the
feasibility condition
\begin{eqnarray}
\sum_{i=1}^{n}A^k_{ir} \leq k/n, \forall  r.
\end{eqnarray}

Under the dynamic DRF mechanism, assume that allocations are {\it
irrevocable}, i.e., $A_{ir}^k\geq A_{ir}^{k-1}$, for every step
$k\geq 2$, every agent $i\leq k-1$, and every resource $r$. At every
step $k$, assume  $\textbf{A}^k$ is non-wasteful, which means that
for every agent $i$ there exists $y\in \mathbb{R}^{+}$ such that for
every resource $r$, $A_{ir}^k=y\cdot d_{ir}$. Let $x^k_i$ be
dominant share of user $i$ at step $k$, which implies
\begin{eqnarray}
A^k_{ir}=x^k_i\cdot d_{ir},  \text{ for } i=1,2,\ldots,k, \text{ and
} r=1,2,\ldots,m.
\end{eqnarray}
At every step $k$, the dynamic DRF mechanism \cite{Kash14} starts
from the current allocation among the present agents $1,\ldots,k$
and keeps allocating resources to agents that have the minimum
dominant share synchronously, until a $k/n$ fraction of at least one
resource is allocated. Formally, at every step $k$, the dominant
share vector $(x_1^k, \ldots, x_k^k)$ of the dynamic DRF allocation
$\textbf{A}^k$ can be obtained by solving the following linear
program:  \begin{eqnarray}\label{x2}
     \left\{ \begin{split}
& \text{Maximize }  M^k\\
& x_i^k \ge M^k,\forall i\leq k;  \\
& x_i^k\ge x_i^{k-1},\forall i \leq k-1; \hspace{2mm}\text{ (irrevocable) } \\
& \sum_{i=1}^k d_{ir}  x_i^k\leq k/n,\forall r. \hspace{2mm}\text{
(capacity constraints) }
        \end{split}\right.
\end{eqnarray}

As shown in \cite{Kash14},  the dynamic DRF mechanism  satisfies
many desired properties. Especially, it satisfies {\it sharing
incentives} (SI) and {\it dynamic Pareto optimality} (DPO). SI means
that, for all steps $k$ and all agents $i\leq k$,  $x_i^k\geq 1/n$,
i.e., when an agent arrives it receives an allocation that it likes
at least as much as an equal split of the resources. DPO means that,
for all steps $k$, there is a resource $r$ such that
$\sum_{i=1}^kd_{ir}x_i^k=k/n$, i.e., it should not be possible to
increase the allocation of an agent without decreasing the
allocation of at least another user, subject to  not allocating more
that $k/n$ fraction of any resource.

\section{Competitive ratios analysis}
In \cite{Kash14}, the authors analyzed the performance of the
dynamic DRF mechanism on real data, for two objectives: the sum of
dominant shares (the maxsum objective) and the minimum dominant
share (the maxmin objective) of the agent present in the system. In
this section, we analyze  the performance of the dynamic DRF
mechanism in the worst-case scenario. For a maximization problem,
the competitive ratio $\rho$ of an online algorithm is the
worst-case ratio between the cost of the solution found by the
online algorithm and the cost of an optimal solution in an offline
setting where all the demands of agents are known \cite{Borodin}.
Clearly, $\rho\in [0,1]$. Similarly, we define the {\it competitive
ratio} of the dynamic DRF mechanism as the worst-case ratio between
the objective value of the dynamic DRF solution
$(x^k_1,\ldots,x^k_k)$
 and the optimal solution $(\tilde{x}^k_1,\ldots,\tilde{x}^k_k)$ of instance $I$ under certain objective function.  Accordingly, the
competitive ratio $CR$ of the dynamic DRF mechanism is defined as
\begin{eqnarray}
CR=\min_{I}\min_{k}\frac{\text{The objective value of
 } (x^k_1,\ldots,x^k_k)}{\text{The objective value of }
(\tilde{x}^k_1,\ldots,\tilde{x}^k_k)}.
\end{eqnarray}

\subsection{The maxsum objective}
 When the objective is the sum of
dominant shares maximization (maxsum, for short), for a given
instance $I$, the optimal solution
$(\dot{x}^k_1,\ldots,\dot{x}^k_k)$ at step $k$ ($\geq 2$) in the
offline setting can be obtained by solving the following program
\begin{eqnarray}\label{x2}
     \left\{ \begin{split}
& \text{Maximize }  \sum_{i=1}^{k}x_i^k \\
 & \sum_{i=1}^k d_{ir}
x_i^k\leq k/n,\forall r.
        \end{split}\right.
\end{eqnarray}
Accordingly, the competitive ratio $CR_1$ of the dynamic DRF
mechanism for the maxsum objective can be defined as
\begin{eqnarray}
CR_1=\min_{I} \min_{k}\frac{\sum_{i=1}^k x^k_i}{\sum_{i=1}^k
\dot{x}^k_i }.
\end{eqnarray}

 \vspace{1mm} \noindent {\bf Theorem~1.} When the objective is the sum of dominant shares maximization,
 the competitive ratio
 of the dynamic DRF mechanism is $1/m$, and the ratio is tight.  \\
{\bf Proof.} Since dynamic DRF mechanism satisfies SI, we have
$x^k_i\geq 1/n$ for every agent $i\leq k$ at step $k$, which implies
that \begin{eqnarray}\sum_{i=1}^{k}x^k_i\geq
\frac{k}{n}.\end{eqnarray} Consider the optimal solution
$(\dot{x}^k_1,\ldots,\dot{x}^k_k)$ obtained from (5). Clearly, at
step $k$,
\begin{eqnarray}\sum_{i: r^*_i=r}\dot{x}^k_i=\sum_{i: r^*_i=r}d_{ir^*_i}\dot{x}^k_i\leq \sum_{i=1}^k d_{ir}
\dot{x}_i^k\leq \frac{k}{n}, \end{eqnarray} for every resource $r$,
following from the fact $d_{ir^*_i}=1$ and the capacity constraint
of (5). It implies that
\begin{eqnarray}\sum_{i=1}^{n}\dot{x}^k_i\leq \sum_{r=1}^{m}\sum_{i: r^*_i=r}\dot{x}^k_i
\leq \frac{mk}{n}, \end{eqnarray} where the first inequality follows
from the fact that each agent has at least one dominant resource.
Thus, following (7) and (9), we have
\begin{eqnarray}\frac{\sum_{i=1}^{k}x^k_i}{\sum_{i=1}^{n}\dot{x}^k_i}\geq \frac{1}{m}, \end{eqnarray}
 {\it i.e.},
the competitive ratio of the dynamic DRF mechanism is at least
$1/m$.

Next, we will prove that the competitive ratio is tight. Consider a
setting with $m$ ($\geq 2$) resources and $n$ ($\gg m$) agents. For
$i=1,2,\ldots,n-m$, the demand vector of agent $i$ is
$\textbf{d}_{i}=(1,1,\ldots,1)$. For agents $i=n-m+1, n-m+2,\ldots,
n$, the demand vectors are $(1,\epsilon, \ldots,\epsilon)$,
$(\epsilon,1, \ldots,\epsilon)$, $\cdots$, $(\epsilon,\epsilon,
\ldots,1)$, respectively, where $\epsilon\rightarrow 0$ is a small
enough number. It is easy to verify that the dynamic DRF mechanism
produces a solution with
\begin{eqnarray}x^n_i=\frac{1}{n-m+1+\epsilon(m-1)}\rightarrow \frac{1}{n-m+1}, \text{ for each agent } i \end{eqnarray}
 at step $n$. The optimal solution will allocate
all resources to last the $m$ agents, obtaining a solution with
\begin{eqnarray}\dot{x}^n_i=\frac{1}{1+\epsilon(m-1)}\rightarrow 1, \text{  for } i=n-m+1,
n-m+2,\ldots, n,  \end{eqnarray} and $\dot{x}^n_i=0$ for other
agents. Thus, the competitive ratio is
\begin{eqnarray}
\frac{\sum_{i=1}^{k}x^n_i}{\sum_{i=1}^{n}\dot{x}^n_i}\rightarrow
\frac{n}{m(n-m+1)}=\frac{1}{m}\frac{1}{1-m/n+1/n}.
\end{eqnarray}
When $n$ is large enough, the ratio approaches $1/m$. Thus, the
theorem holds. \q
\subsection{The maxmin objective}

When the objective is minimum dominant share maximization  (maxmin,
for short), the optimal solution
$(\ddot{x}^k_1,\ldots,\ddot{x}^k_k)$ at step $k$ ($\geq 2$) in the
offline setting can be obtained by solving the following program
\begin{eqnarray}\label{x2}
     \left\{ \begin{split}
& \text{Maximize }  \min_{i} x_i^k \\
 & \sum_{i=1}^k d_{ir}
x_i^k\leq k/n,\forall r.
        \end{split}\right.
\end{eqnarray}
Actually, $(\ddot{x}^k_1,\ldots,\ddot{x}^k_k)$ is a DRF solution
\cite{Ghodsi11, Parkes15}, where the  dominant shares of all agents
are equal. Formally, for a given instance $I$, at every step $k$,
 $(\ddot{x}^k_1,\ldots,\ddot{x}^k_k)$ is obtained by
\begin{eqnarray}
\ddot{x}^k_1=\cdots=\ddot{x}^k_k=\min_r\frac{k/n}{\sum_{i=1}^{k}d_{ir}},
\end{eqnarray}
following from \cite{Gutman12,Parkes15}.

Therefore, the competitive ratio $CR_2$ of the dynamic DRF mechanism
for the maxmin objective can be defined as
\begin{eqnarray}
CR_2=\min_{I} \min_{k}\frac{\min_{i} x^k_i}{\min_{i}
\ddot{x}^k_i}=\min_{I} \min_{k}\frac{\min_{i} x^k_i}{
\ddot{x}^k_k}=\min_{I} \min_{k}\frac{x^k_k}{\ddot{x}^k_k},
\end{eqnarray}
where  the last equality follows from the fact  $\min_{i}
x^k_i=x^k_k$, which can be obtained by Lemma 2 in \cite{Kash14}.

 \vspace{1mm} \noindent {\bf Theorem~2.} When the objective is minimum dominant share maximization,
 the competitive ratio
 of the dynamic DRF mechanism is $1/m$. Moreover, no mechanism satisfying DPO can do
 better than $1/(m-1)$. \\
{\bf Proof.}  At every
 step $k\in \{2,\ldots,n\}$,  since the dynamic DRF mechanism  satisfies the SI property, we
have  \begin{eqnarray}x^k_k\geq \frac{1}{n}. \end{eqnarray} By the
pigeonhole principle, there exists a source which is the dominant
resource for at least $\lceil k/m\rceil$ agents. It implies that the
DRF solution $(\ddot{x}^k_1,\ldots,\ddot{x}^k_k)$ satisfies
\begin{eqnarray}\ddot{x}^k_k\leq \frac{k/n}{\lceil k/m\rceil}. \end{eqnarray} Thus, the
competitive ratio
 of  the dynamic DRF mechanism  satisfies
 \begin{eqnarray}
 CR_2=\frac{x^k_k}{\ddot{x}^k_k}\geq \frac{\lceil k/m\rceil}{k}\geq \frac{1}{m}.
 \end{eqnarray}

Consider a setting with $m$ ($> 2$) resources and $n=m^2+1$ agents.
For $i=1,2,\ldots,m^2$, the demand vector of agent $i$ is defined as
\begin{eqnarray}
\textbf{d}_{i}= \left\{ \begin{split}
& (1,\epsilon,\ldots,\epsilon), \text{ if } i\equiv 1 \text {  (mod } m)\\
& (\epsilon,1,\ldots,\epsilon), \text{ if }  i\equiv 2 \text {  (mod } m)  \\
& \hspace{20mm} \cdots \\
& (\epsilon,\epsilon,\ldots,1), \text{ if }  i\equiv 0 \text { (mod
} m)
\end{split}, \right.
\end{eqnarray}
where $\epsilon\rightarrow 0$ is a small enough number. The demand
vector of agent $n=m^2+1$ is $\textbf{d}_{n}=(1,1,\ldots,1)$. At
step $k=m^2$, the dynamic  DRF solution is \begin{eqnarray}
x^k_i=\frac{m^2}{[m+\epsilon(m^2-m)](m^2+1)} \rightarrow
\frac{m}{m^2+1}, \text{ for } i=1,2,\ldots,k,
\end{eqnarray}following from the assumption of $\epsilon$.
Actually, after the first $m^2$ steps, at least $m^2/(m^2+1)$ share
of at least one resource $r^*$ must be exhausted for any dynamic
mechanism satisfying the DPO property. It implies that at most
$1/(m^2+1)$ share of resource $r^*$ is left for the last agent $n$.
Hence,   \begin{eqnarray}x^n_n\leq \frac{1}{m^2+1}, \end{eqnarray}
for any dynamic mechanism satisfying DPO, while the DRF solution
$(\ddot{x}^n_1,\ldots,\ddot{x}^n_n)$ satisfies
\begin{eqnarray}\ddot{x}^n_i= \frac{1}{m+1+\epsilon(m^2-m)} \rightarrow
\frac{1}{m+1}, \text{ for } i=1,2,\ldots,n.\end{eqnarray} It implies
that, at step $k=n$,  the competitive ratio
 of any dynamic mechanism satisfying
DPO including the dynamic DRF mechanism is at most
\begin{eqnarray} \frac{  x^n_n}{\ddot{x}^n_n}\rightarrow
\frac{m+1}{m^2+1}\leq  \frac{1}{m-1}.\end{eqnarray} Thus, the
theorem holds. \q

\section{A linear-time optimal algorithm}
Since dynamic DRF is almost optimal as proved in the last section,
it is desired to design an efficient algorithm to find an optimal
solution for the dynamic DRF mechanism. Although the  water-filling
algorithm can produce a dynamic DRF solution \cite{Kash14}, the
running time is pseudo-polynomial \cite{Gutman12}. Also, we can
compute a dynamic DRF solution by solving the linear program (1).
However, it is not a strongly polynomial-time algorithm. In this
section, we will design a linear-time algorithm to find a dynamic
DRF solution.   %At every step
%$k$, by modifying the linear-time algorithms for computing a DRF
% solution in \cite{Gutman12,Parkes15} slightly, we can obtain a
%variant of water-filling algorithm with running time $O(k^2)$.
%Actually, we can do even better.
In the proof below, $M^k$ and $x^{k}_i$ refer to the optimal
solution of (1) in step $k$. The following two lemmas in
\cite{Kash14} are very useful for designing the faster algorithm.

 \vspace{1mm} \noindent {\bf Lemma~1.}   At any
 step $k\in \{1,\ldots,n\}$, it holds that $x^{k}_i=\max \{M^k, x^{k-1}_i\}$ for all agents $i\leq
 k$.

 \vspace{1mm}  \noindent {\bf Lemma~2.}   At any
 step $k\in \{1,\ldots,n\}$,  for all agents $i,j$ such that $i<j$,
  it holds that $x^{k}_i\geq x^{k}_j$.

%  \vspace{1mm} \noindent {\bf Lemma~3.}   At any
% step $k\in \{1,\ldots,n\}$,  if $x^{k}_i> x^{k}_j$ for some agents $i,j\leq k$, then
% $i<j$ and   $x^{k}_i= x^{j-1}_i$.

 \vspace{1mm} \noindent {\bf Theorem~3.} At any
 step $k\in \{2,\ldots,n\}$,  a dynamic DRF solution can be found within $O(k)$ time. \\
{\bf Proof.} Consider an agent $j$. By Lemma 1, we have
$x^{k}_j=\max \{M^k, x^{k-1}_j\}$. If $x^{k}_j=x^{k-1}_j>M^k$, by
Lemma 2, for all agents $i\leq j$, we have $x^{k-1}_i\geq
x^{k-1}_j>M^k$, which implies that $ x^{k}_i=\max \{M^k,
x^{k-1}_i\}=x^{k-1}_i$. If $x^{k}_j=M^k>x^{k-1}_j$, by Lemma 2, for
all agents $i\geq j$,  we have $x^{k-1}_i\leq x^{k-1}_j<M^k$, which
implies that $ x^{k}_i=\max \{M^k, x^{k-1}_i\}=M^k$. Therefore, at
any step $k\geq 2$, there is an agent $\tau \leq k$ such that
\begin{eqnarray}
\left\{\begin{split}
    & x^{k}_i=x^{k-1}_i>M^k,   \text{ for  } i<\tau;\\
&x^{k}_i= M^k\geq x^{k-1}_{i}, \text{ for  } \tau\leq i\leq k.
\end{split}\right.
\end{eqnarray}
Thus, if we know $\tau$, $M^k$ can be obtained by solving the
following linear program
\begin{eqnarray*}
  \left\{\begin{split}   &   \text{Maximize }  M^k\\
&  \sum_{i: \tau\leq i \leq k}  d_{ir}  M^k+\sum_{i:   i <
\tau}d_{ir}  x^{k-1}_i \leq \frac{k}{n}, \text{ for  }
r=1,2,\ldots,m.\end{split}\right.
\end{eqnarray*}
As pointed in \cite{Parkes15},  this linear program can be rewritten
as
  \begin{eqnarray*}
M^k=\min_r\frac{k/n-\sum_{i:   i < \tau}d_{ir} x^{k-1}_i}{\sum_{i:
\tau\leq i \leq k}d_{ir}}.
\end{eqnarray*}

We are now ready to describe our linear-time algorithm. Our main
idea is to find $\tau$  by using a bisection method. At any step
$k\geq 2$, consider the agent $l=\lceil (1+k)/2 \rceil$. Let
\begin{eqnarray}
\left\{\begin{split}
    & \tilde{x}^{k}_i=x^{k-1}_i,   \text{ for  } i<l;\\
& \tilde{x}^{k}_i=x^{k-1}_{l}, \text{ for  } l\leq i\leq k,
\end{split}\right.
\end{eqnarray}
For convenience, let
\begin{eqnarray}
\left\{\begin{split}
    & \alpha_r=\sum_{i:i<l}  d_{ir}x_i^{k-1},  \forall r;\\
& \beta_r=\sum_{i:l\leq i\leq k}d_{ir}, \forall r.
\end{split}\right.
\end{eqnarray}
  Clearly, if $\alpha_r+x^{k-1}_{l}\beta_r\leq k/n$ for every
  resource $r$, i.e., $(\tilde{x}^{k}_1,\ldots,\tilde{x}^{k}_k)$ satisfies the capacity
constraints in (1), we have $M^k\geq x^{k-1}_{l}$ and $l\geq \tau$.
Otherwise,
   we have $M^k< x^{k-1}_{l}$ and $l< \tau$. We distinguish the following two cases:

{\bf Case 1}. $l\geq \tau$. For every agent $i$ satisfying $l\leq
i\leq k$, we have $x_i^k=M^k$, as $i\geq l \geq \tau$. Let ${\cal
AI}=\{i: l\leq i\leq k\}$ be set of known agents with identical
dominant share in the optimal solution $(x^{k}_1,\ldots,x^{k}_k)$.
Next, consider the agent $\lceil(1+l)/2\rceil$ as before.

{\bf Case 2}. $l< \tau $. For every agent $i$ satisfying $i<l$, we
have $x_i^k=x^{k-1}_i$, as $i< l < \tau$. Let ${\cal AS}=\{i: i<l\}$
be set of known agents with same dominant share as in step $k-1$ in
the optimal solution $(x^{k}_1,\ldots,x^{k}_k)$. Next, consider the
agent $\lceil(l+k)/2\rceil$ as before.

 At every step $k$, the number of unclassified agents
in $\{i: i\notin {\cal AI}, i\notin{\cal AS}\}$ is
  reduced to half. Finally, all the agents are divided into two subsets ${\cal AI}$ and ${\cal
  AS}$,
  and we will find the $\tau$ and the optimal solution $(x^k_1,\ldots,x^k_k)$.
 Clearly, the running time of deciding whether $l\geq \tau$ at each iteration
  is linear in the number of unclassified agents. Thus,  the total running time is
  $O(k+k/2+k/{2^2}+\cdots+1)=O(k)$, where $m$ is seen as a
constant. \q

The complete algorithm is given as   {\sc Linear-time dynamic DRF
algorithm} in Appendix.

\section{Conclusion and Future Work}
 We have analyzed the competitive ratio of the
dynamic DRF mechanism, which shows that the dynamic DRF mechanism is
a nearly optimal mechanism satisfying DPO for the maxmin objective.
We have described a non-trivial polynomial-time algorithm to find a
dynamic DRF allocation, whose running time is linear in the number
of present agents at every step, improving the result in
\cite{Kash14}.

 Note that another fair allocation mechanism, called {\it cautious LP}, is proposed in
  \cite{Kash14}.  Cautious LP
achieves near optimal maxmin value at the last step. However, since
cautious LP violates the DPO property and allocates too many
resources at the last several steps, it is unfair to compare
cautious LP with dynamic DRF for the maxmin objective. It is
interesting to analyze the competitive ratio of the cautious LP
mechanism under different objectives. Since solving the linear
program takes too much time, it is challenging to develop a
combinatorial algorithm to find a cautious LP solution as in Section
4.

%\section*{Conflict of Interests} The authors declare that there is no conflict
%of interests regarding the publication of this paper.

\section*{Acknowledgment}
The work is supported in part by the National Natural Science
Foundation of China [Nos. 11301466, 61170222],  and the Natural
Science Foundation of Yunnan Province of China [No. 2014FB114].

\section*{Appendix}
\begin{center}
\begin{tabular}[t]{l}
\hline  {\sc Linear-time dynamic DRF
algorithm}  \\
\hline
\hspace{1mm} 1:   {\bf Data}: Demand ${\bf d}_i$, $1\leq i\leq k$\\
\hspace{1mm} 2:  {\bf Result}: Allocation ${\bf A}^k$ at each step $k$\\
\hspace{1mm} 3:   $x^{1}_1\leftarrow   1/n$, $A^1_{1r}\leftarrow x^1_1\cdot d_{1r}$, $\forall r$;\\
\hspace{1mm} 4:   $k\leftarrow 2$;\\
\hspace{1mm} 5:  \textbf{while} $k\leq n$ \textbf{do}\\
\hspace{1mm} 6:  \hspace{3mm} \textbf{if} $\sum_{i=1}^{k} d_{ir}x^{k-1}_1\leq k/n$, $\forall r$, \textbf{do}  \\
\hspace{1mm} 7:  \hspace{3mm}  $\alpha_r\leftarrow  0$, $\beta_r\leftarrow  \sum_{i=1}^{k} d_{ir}$, $\forall r$;\\
\hspace{1mm} 8: \hspace{3mm}  \textbf{else, do}\\
\hspace{1mm}  9: \hspace{3mm} $LB\leftarrow 1$, $UB\leftarrow k$, $\tau\leftarrow \lceil(LB+UB)/2\rceil$;\\
10: \hspace{2mm} $\alpha_r\leftarrow \sum_{i=1}^{\tau-1} d_{ir}x^{k-1}_i$, $\beta_r\leftarrow \sum_{i=\tau}^{k}d_{ir}$, $\forall r$;\\
11: \hspace{5mm} \textbf{while} $UB-LB>1$, \textbf{do} \\
12: \hspace{8mm} \textbf{if} $\alpha_r+\beta_rx^{k-1}_{\tau}\leq
k/n$, $\forall r$, \textbf{do} \\
13:  \hspace{8mm}  $LB\leftarrow LB$, $UB\leftarrow \tau$,  $\tau\leftarrow \lceil(LB+UB)/2\rceil$;\\
14:  \hspace{8mm}  $\alpha_r\leftarrow \alpha_r-\sum_{i=\tau}^{UB-1} d_{ir}x^{k-1}_i$, $\beta_r\leftarrow \beta_r+\sum_{i=\tau}^{UB-1} d_{ir}$;\\
15:  \hspace{8mm} \textbf{else, do}\\
16:  \hspace{8mm} $LB\leftarrow \tau$, $UB\leftarrow UB$, $\tau\leftarrow \lceil(LB+UB)/2\rceil$;\\
17:  \hspace{8mm}  $\alpha_r\leftarrow \alpha_r+\sum_{i=LB}^{\tau-1} d_{ir}x^{k-1}_i$, $\beta_r\leftarrow \beta_r-\sum_{i=LB}^{\tau-1} d_{ir}$;\\
18:  \hspace{8mm} \textbf{end if}; \\
19:  \hspace{5mm} \textbf{end while}; \\
20:  \hspace{2mm} \textbf{end if}; \\
21:  $M^k\leftarrow \min_r (k/n-\alpha_r)/\beta_r$;\\
22:  $x^k_i\leftarrow \max (x^{k-1}_i,M^k), \forall i\leq k;$\\
23:  $A^k_{ir}\leftarrow x^k_i\cdot d_{ir}, \forall i\leq k;$\\
24:  $k\leftarrow k+1$;\\
25:  \textbf{end while} \\
\hline
\end{tabular}
\end{center}


\begin{thebibliography}{}
%\bibitem{Bertsimas11} D. Bertsimas, V. F. Farias, and N. Trichakis, The price of
% fairness, Operations Research 59(1), pp. 17-31, 2011.
%\bibitem{Bhattacharya13} A.A. Bhattacharya, D. Culler, E. Friedman, A. Ghodsi,
%S. Shenker, and I. Stoica, Hierarchical scheduling for diverse
%datacenter workloads. In Proceedings of the 4th Annual Symposium on
%Cloud Computing, SOCC'13, Article No. 4, 2013.

%\bibitem{Blum} M. Blum, R.W. Floyd, V. Pratt, R.R. Rivest, R.E. Tarjan, Time bounds
%for selection, Journal of Computer and System Sciences 7(4),
%448-461, 1973.


\bibitem{Borodin} A. Borodin,  and R. El-Yaniv, Online computation and competitive
analysis, Cambridge University, 1998.


%\bibitem{Bonald14} T. Bonald, J. Roberts, Enhanced cluster computing
% performance through proportional fairness, Performance Evaluation 79, 134-145, 2014.


% \bibitem{Bonald142} T. Bonald, J. Roberts, Multi-resource fairness: Objectives, algorithms and performance,
% arXiv:1410.0782, 2014.

%\bibitem{Dolev12} D. Dolev, D. G. Feitelson, J. Y. Halpern, R. Kupferman, and N.
% Linial, No justified complaints: on fair sharing of multiple
%resources. In Proceedings of the 3rd Innovations in Theoretical
%Computer Science Conference, ITCS'12, pp. 68-75, 2012.


%\bibitem{Friedman14} E. Friedman, A. Ghodsi, C-A. Psomas,
% Strategyproof allocation of discrete jobs on multiple machines, in
%  Proceedings of the fifteenth ACM conference on Economics and
%  computation, pp. 529-546, 2014.

\bibitem{Ghodsi11}A. Ghodsi, M. Zaharia, B. Hindman, A. Konwinski, S. Shenker, and
I. Stoica, Dominant resource fairness: fair allocation of multiple
resource types. In Proceedings of the 8th USENIX Conference on
Networked Systems Design and Implementation, NSDI'11, pp. 24-24,
2011.

\bibitem{Gutman12} A. Gutman and N. Nisan, Fair allocation without trade. In
Proceedings of the 11th International Conference on Autonomous
Agents and Multiagent Systems, AAMAS'12, pp. 719-728, 2012.

\bibitem{Kash14} I. Kash, A. D. Procaccia, N. Shah, No agent left behind:
dynamic fair division of multiple resources, Journal of Articial
Intelligence Research 51, pp. 579-603, 2014.


%\bibitem{Kelly98} F.P. Kelly, A.K. Maulloo, D.K.H. Tan, Rate control for communication
%networks: shadow prices, proportional fairness and stability,
% Journal of the Operational Research Society 49 (3), pp. 237-252,
% 1998.

%\bibitem{Li14} W. Li, X. Liu, X. Zhang, and X. Zhang, Multi-resource fair
%allocation with bounded number of tasks in cloud computing systems,
%arXiv:1410.1255, 2014.



\bibitem{Parkes15} D.C. Parkes, A.D. Procaccia,
and N. Shah, Beyond dominant resource fairness: extensions,
limitations, and indivisibilities, ACM Transactions on Economics and
Computation 3(1) (2015), Article no. 3.



% \bibitem{Procaccia13} A.D. Procaccia, Cake cutting: not just child¡¯s play, Communications of the ACM, 2013.



%\bibitem{Psomas13} C.-A. Psomas and J. Schwartz, Beyond beyond
%dominant resource fairness: indivisible resource allocation in
%clusters, Tech Report Berkeley, 2013.



%\bibitem{Megiddo} N. Megiddo, Optimal flows in networks with multiple sources and
% sinks, Mathematical Programming 7(3), pp. 97-107, 1974.



\bibitem{Wang14} W. Wang,  B. Liang, and B. Li,
Multi-resource fair allocation in heterogeneous cloud computing
systems, IEEE Transactions on Parallel and Distributed Systems
26(10), 2822-2835, 2015.

%\bibitem{Wilkes} J. Wilkes and C. Reiss, google ClusterData2011\_2,

%https://code.google.com/p/googleclusterdata/.

%\bibitem{Varian74} H. Varian, Equity, envy, and efficiency, Journal of Economic
%Theory 9(1), pp. 63-91, 1974.

\bibitem{Wong13} C. Joe-Wong, S. Sen, T. Lan, and M. Chiang, Multi-resource
allocation: Fairness-efficiency tradeoffs in a unifying framework,
IEEE/ACM Transactions on Networking 21(6), pp. 1785-1798, 2013.

\bibitem{Zarchy} D. Zarchy, D. Hay, M. Schapira, Capturing resource tradeoffs in fair
multi-resource allocation, Infocom 2015, pp. 1062-1070, Hong Kong,
 2015.

%\bibitem{Zahedi14} S. M. Zahedi, and  B. C. Lee,  REF: Resource elasticity
%fairness with sharing incentives for multiprocessors, In Proceedings
%of the 19th International Conference on Architectural Support for
%Programming Languages and Operating Systems (ASPLOS), pp. 145-160,
%2014.


%\bibitem{Zeldes13} Y. Zeldes and D. G. Feitelson, On-line fair allocations based on
%bottlenecks and global priorities. In Proceedings of the 4th
%ACM/SPEC International Conference on Performance Engineering, ICPE
%¡¯13, pp. 229-240, 2013.




\end{thebibliography}
\end{document}